\documentclass[twocolumns]{wlscirep}  
                           
\usepackage{amsmath}
\usepackage{amsthm}
\usepackage{dcolumn}
\usepackage{bm}
\usepackage{amssymb}
\usepackage{graphicx}
\usepackage{epstopdf}
\usepackage[flushleft]{threeparttable}
\usepackage{textcomp}
\usepackage{color}
\usepackage{tabularx}
\usepackage[normalem]{ulem}

\newtheorem{definition}{Definition}
\newtheorem{proposition}{Proposition}

\usepackage{lipsum}

\let\OLDthebibliography\thebibliography
\renewcommand\thebibliography[1]{
  \OLDthebibliography{#1}
  \setlength{\parskip}{0pt}
  \setlength{\itemsep}{0pt plus 0.3ex}
}

\title{On neighbourhood degree sequences of complex networks}
\author[1,*]{Keith M. Smith}
\affil[1]{Usher Institute of Population Health Science and Informatics, University of Edinburgh, 9 BioQuarter, Little France, Edinburgh, EH16 4UX, UK}%
\affil[*]{k.smith@ed.ac.uk}

\begin{abstract}
Network topology is a fundamental aspect of network science that allows us to gather insights into the complicated relational architectures of the world we inhabit. We provide a first specific study of neighbourhood degree sequences in complex networks. We consider how to explicitly characterise important physical concepts such as similarity, heterogeneity and organisation in these sequences, as well as updating the notion of hierarchical complexity to reflect previously unnoticed organisational principles. We also point out that neighbourhood degree sequences are related to a powerful subtree kernel for unlabelled graph classification. We study these newly defined sequence properties in a comprehensive array of graph models and over 200 real-world networks. We find that these indices are neither highly correlated with each other nor with classical network indices. Importantly, the sequences of a wide variety of real world networks are found to have greater similarity and organisation than is expected for networks of their given degree distributions. Notably, while biological, social and technological networks all showed consistently large neighbourhood similarity and organisation, hierarchical complexity was not a consistent feature of real world networks. Neighbourhood degree sequences are an interesting tool for describing unique and important characteristics of complex networks.
\end{abstract}

\begin{document}
\maketitle

\section{Introduction}	
\begin{figure}[!tb]
	\centering
	\includegraphics[trim = 0 80 0 50,clip,scale=.4]{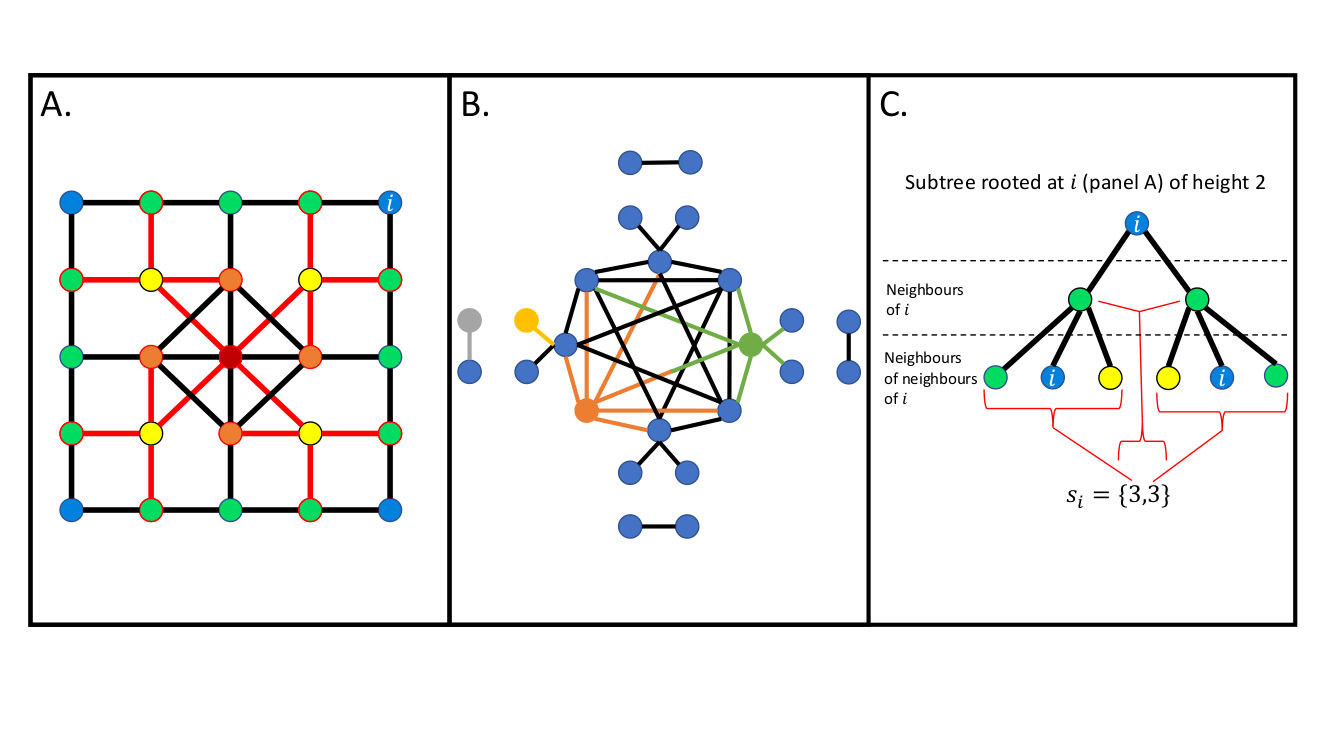}
	\caption{A. How can we efficiently capture the organisation of this graph mathematically without reference to node placements on the plane? We can note that the neighbourhoods of nodes of a given degree are equivalent with respect to the degrees of nodes they connect to-- e.g. all yellow nodes (degree 4) connect to the same number of green (degree 3), orange (degree 5) and red (degree 8) nodes. Thus neighbourhood degree sequences appear as a promising avenue. B. Illustration of a multi-ordered degree graph whose equal-degree nodes are organised into two distinct classes with different degree sequences. C. Illustration of a subtree of height 2 for node $i$ in panel A. The number of nodes at height 1 is the degree of $i$, while for a node at height 1, its degree is the the number of nodes at height 2 extending from it, all captured by $i$'s neighbourhood degree sequence, $s_{i}$.}
	\label{NhG}
\end{figure}

Contemplating the roles of components in natural and man-made systems, we begin to realise their diversity. Take for example, the structure of an organisation. At face value, employees are assigned titles and pay-scales which place the workforce in a convenient hierarchy with each level comprising of equivalencies based on the competitive value of the work done. However, in large and multifaceted organisations the work done is often highly variable and it is beneficial to have employees with a diverse range of skills and talents interacting in different ways. Network science provides a natural framework to understand relationship patterns of such complex systems and we shall here formulate and study hierarchical equivalency in terms of neighbourhood degree sequences of complex networks. Fig \ref{NhG}.A provides an illustration of how neighbourhood degree sequences intuitively help to understand global hierarchical patterns.

The distribution of connections among nodes in complex networks, known as the degree distribution, is a key consideration of its topology. Predated by the study of degree sequences \cite{Bollobas1981}, interest in degree distributions arose from the study of real-world networks, where it was noted that they approximated various statistical distributions with heavy tails \cite{Strogatz2001}, being particularly driven by the prevalence of strong hubs in real-world networks which are not present, for example, in random graphs \cite{Erdos1959}, random geometric graphs \cite{Dall2002} and small-world models \cite{Watts1998}. Pertinent random null models, called configuration models, have since been developed in which the degree distribution is fixed, allowing unbiased random controls for studying network topologies \cite{Newman2001, Maslov2002}.

Although often explicitly mentioned with regard to real-world networks, what is meant by concepts such as organisation and complexity has largely been left to intuition. In seeking to understand the complexity of real world networks, Smith \& Escudero \cite{Smith2017a} recently proposed to look at neighbourhood degree sequences. For a given node, its neighbourhood degree sequence was defined as the ordered degrees of nodes in its neighbourhood. This was based on observations that ordered networks such as regular networks, quasi-star networks, grid networks and highly patterned networks shared the common feature of highly homogeneous neighbourhood degree sequences for nodes of the same degree. Conceptualising the degree distribution as a hierarchy of nodes, they proposed an index called hierarchical complexity to characterise the heterogeneity of hierarchically equivalent (i.e. same degree) nodes. Note, the term `hierarchy' in networks is also associated with the scaling of community structure \cite{Ravasz2003, Kaiser2010}. Here, it is used-- in the more lexically familiar sense-- with respect to levels of importance, where nodes of higher degree are often considered of higher importance in the network topology \cite{Barthelemy2004}. Hierarchical complexity was developed in the context of electroencephalogram functional connectivity, which, in contrast to ordered and random systems, was found to have inordinately high levels of heterogeneity amongst its neighbourhood degree sequences \cite{Smith2017a}. This concept has since been utilised to help understand how best to binarise EEG functional connectivity for topological analysis \cite{Smith2017c} and has been validated in structural MRI networks \cite{Smith2019}. However, the prevalence of such topology amongst complex networks in general is unknown. In pure mathematics, Barrus \& Donovan independently initiated study of neighbourhood degree lists as a topological invariant more refined than both the degree sequence and joint degree graph matrix \cite{Barrus2018}, while Nishimura \& Subramanya proposed to study neighbourhood degree lists for the combinatorial problem of changing a graph into one with given neighbourhood degrees \cite{Nishimura2017}.

That is as far as has been done with neighbourhood degree sequences to date. Yet, the intriguing insights provided by hierarchical complexity in brain networks makes a broader study of neighbourhood degree sequences across a broader range of domains worthwhile. This work comes after work done involving neighbouring degrees and centralities such as the eigenvector centrality, a centrality index which is larger depending on the centralities of the nodes a node is connected to \cite{Bonacich1972}; assortativity, an index of degree-degree correlation between connected nodes \cite{Newman2002}; and network entropy, a measure of edgewise node degree eccentricity \cite{Sole2004}. Neighbourhood degree sequences, however, are a completely separate consideration of networks. Most notably, rather than comparing nodes which are connected to each other, we compare nodes which have the same degree, irrespective of whether they are connected or not, regarding such nodes as hierarchically equivalent within the network topology.

In this study a number of ways to analyse neighbourhood degree sequences are proposed. Notably, indices of node heterogeneity and neighbourhood similarity are introduced. We also consider a new notion of multi-orderedness in a network. This is based on the observation that nodes of a given degree in an ordered network may have several distinct neighbourhood degree sequences. This gives rise to another index defined as neighbourhood organisation which measures the extent to which such multi-orderedness is present in the network. We then show that the existence of multi-ordered degrees can artificially raise the network's hierarchical complexity. Thus, we utilise the formulation of neighbourhood organisation to provided a version of hierarchical complexity which corrects for multi-ordered degrees. We also described how neighbourhood degree sequences have clear links with powerful and efficient subtree kernels for graph classification. The proposed indices are then applied to a range of network models and compared with existing classical network indices, the aim of which is to ascertain to what extent these indices explain unique topological properties in complex networks. They are also applied to 215 real world networks from various disciplines of study in order to assess the characteristics of neighbourhood degree sequences in the world around us and the insights these new indices offer.

\section{Neighbourhood Degree Sequences}
For $k_{i}$ the degree of node $i$, the neighbourhood degree sequence, $s_{i}$, of node $i$ is 
\begin{equation}
s_{i} = \{k^{i}_{1},k^{i}_{2},\dots,k^{i}_{k_{i}}\},
\end{equation}
where the $k_{j}^{i}$s are the degrees of the nodes to which $i$ is connected and such that $k_{1}^{i}\leq k_{2}^{i}\leq \dots \leq k_{k_{i}}^{i}$. For example, the graph in Fig \ref{NhG}.A has four degree 4 nodes (yellow) all with neighbourhood degree sequence $\{3,3,5,8\}$ and four degree 5 nodes (orange) all with neighbourhood degree sequence $\{3,4,5,5,8\}$. In the following we shall consider a number of ways to study these sequences.

\subsection{Node Heterogeneity}
One way to characterise neighbourhood degree sequences would be to employ the same methods to characterise degree distributions and then average over all nodes. As a pertinent example of this, a common index of graph heterogeneity is the degree variance $v = \text{var}(k)$ \cite{Snijders1981}. We can then define node heterogeneity, $V_{n}$, as the average variance of neighbourhood degree sequences of a graph for all nodes of degree greater than 1:
\begin{equation}\label{Vn}
V_{n}(G) = \frac{1}{n}\sum_{i \text{ s.t. } k_{i}>1}\text{var}(s_{i}).
\end{equation}
Of course, it is then interesting to understand how average node heterogeneity compares to graph heterogeneity, i.e. comparing local and global heterogeneities of a graph. To do this we can simply divide \eqref{Vn} by $v$, giving
\begin{equation}\label{Vn}
\hat{V}_{n}(G) = \frac{1}{n\text{var}(k)}\sum_{i \text{ s.t. } k_{i}>1}\text{var}(s_{i}).
\end{equation}
High values of this measure tell us that nodes tend to be connected to nodes of homogeneous degrees, given the degree distribution, and low values tell us the opposite. Specifically, if this value is below 1, the degree variance within the neighbourhoods is on average less than the global degree variance, indicating that the nodes have more homogeneous neighbourhood degrees. It is worth highlighting the distinction between this and assortativity, which seeks to measure the similarity of degrees of connected nodes. Node heterogeneity is a measure of the similarity of the degrees of all neighbouring nodes, irrespective of the degree of the node itself.

Note that $v$ is clearly minimal for regular graphs and is known to be maximal for quasi-star and quasi-complete graphs for any given number of nodes and edges \cite{Bell1992}. On the other hand $V_{n}$ is zero for regular graphs but is also small for quasi-star and quasi-complete graphs. For instance, the star graph consists of one node connected to all other nodes and no other edges. Thus it has one $n-1$ degree node with degree sequence $\{1,1,\dots,1\}$ and $n-1$ 1 degree nodes with degree sequence $\{n-1\}$. Clearly, these all have zero variance, giving $V_{n} = 0$ for the star graph. This is interesting because, while some believe star graphs should have maximum heterogeneity \cite{Estrada2010}, $V_{n}$ points at a possible different view. The degree distribution of a star graph is just 1 node away from being completely regular-- take the dominant node out and you have an empty graph (redundantly regular). Heterogeneity could perhaps be alternatively formulated in the sense that removing or adding nodes does not relegate the graph to being regular.

\subsection{Neighbourhood similarity}
The other way of characterising neighbourhood degree sequences we shall consider is to compare all neighbourhood degree sequences of equal length. Indeed, this is the perspective employed to formulate hierarchical complexity, looking at the element-wise variance of equal-length neighbourhood degree sequences. Another, fairly more simple characteristic can be posed by considering the number of nodes in the network whose neighbourhood degree sequence matches that of another node in the graph. We call this neighbourhood similarity (reflecting the concept of geometric similarity) and, using the Kronecker delta function $\delta(x,y)$ which is 1 if $x = y$ and 0 otherwise, write

\begin{equation}
S(G) = \frac{\sum_{i=1}^{n}(1-\delta(\sum_{j=1}^{n}\delta(s_{i},s_{j}),0))}{n}.
\end{equation}
Notice, this uses the $\delta$ function twice. The first time is to find the number of matching neighbourhood degree sequences for node $i$. The second delta is used to determine if there are \emph{any} matching sequences, i.e. seeing if the sum of the first $\delta$s is different from 0. Since this is a negation ($\delta$ returns 0 if there are any matches), we then have to subtract the answer from 1 to provide the answer to whether any match exists for node $i$. Summing over all $i$ and dividing by $n$ provides the proportion of nodes which have at least one matching neighbourhood degree sequence. It is clear that $0\leq S\leq 1$ for all graphs, since it concerns a fraction of the network nodes. It certainly attains 1 for regular graphs. However, we prove the following result with respect to graph symmetry on the plane, establishing the link between neighbourhood similarity and graph symmetry.

\begin{proposition}
Let $G$ be a graph which can be arranged on the plane such that $G$ has mirror or rotational symmetry whose axis does not pivot on any node. Then $S(G) = 1$.
\begin{proof}
Let $s_{i}$ be a neighbourhood degree sequence for general node $i$. Then the node, $j$, in the position symmetric to $i$ with respect to the axis of symmetry has neighbourhood degree sequence $s_{j}$ and has the same degree as $i$. Further, each node in the neighbourhood of $i$, $p_{i}$, also has a node in position symmetric to $p_{i}$ with respect to the axis of symmetry, $p_{j}$, and these nodes are connected to $j$ and such that $k_{p_{i}} =k_{p_{j}}$, by symmetry. Thus $s_{i} = s_{j}$ and since $s_{i}$ was arbitrary and no nodes lie on the axis of symmetry itself, $S(G) = 1$, as required.
\end{proof}
\end{proposition}
Thus, neighbourhood similarity of a graph is indeed related to the planar symmetry of a graph. That being said, the opposite is not true-- not all values $S(G) = 1$ are attained by planar symmetric graphs, as can be quickly seen by regarding non-symmetric regular graphs such as the Frucht graph \cite{Frucht1939}.

\subsection{Hierarchical complexity: oversights of multi-ordered degree graphs}
Hierarchical complexity is an index developed with the aim to be low for all highly ordered graphs and graphs with simple generative mechanisms. Simple in the sense that one needs only a few rules to compute the graph such as in random graphs (edges exist with uniformly random probabilities) or random geometric graphs (nodes are randomly sampled on a n-D Euclidean space and then connected based on distances in the space). In this sense, one can describe precisely how one can expect the graph and subsamples of the graph to behave. On the other hand, attempts to model real world networks indicates that a larger and more a complicated set of rules would be required to generate complex network-like topologies where subsamples of the graph (such as node neighbourhoods) would be less likely to show similar behaviours \cite{Smith2019}. The hypothesis is that nodes of a given degree in highly ordered graphs play equivalent roles in the topology, which implies that they have the same or similar neighbourhood degree sequences. However, what fails to be taken account of in its formulation is the possibility to have a high degree of order in which nodes of a given degree can be split into different groups of identical sequences. For example, Fig \ref{NhG}.B shows a graph with degree 1 and 6 nodes. The six-degree nodes fall into one of two sequences $\{1,1,6,6,6,6\}$ and $\{6,6,6,6,6,6\}$, as illustrated by the green and orange nodes, respectively. One-degree nodes are connected to either one- or six-degree nodes, as illustrated by the grey and yellow nodes, respectively. We call such a graph here a \emph{multi-ordered degree graph}.

\begin{definition}
Let $q_{p}$ be the number of all $p$-length neighbourhood degree sequences and $\sigma_{p} = \{s_{i}\}_{k_{i} = p}$ be the set of (unique) $p$-length neighbourhood degree sequences. Then $p$ is a \emph{multi-ordered degree} of the graph if $1<|\sigma_{p}|<<q_{p}$. A graph for which $1<|\sigma_{p}|<<q_{p}$ or, otherwise, $|\sigma_{p}|= 1$ for all $p$ is called a \emph{multi-ordered degree graph}.
\end{definition}

\subsection{Neighbourhood organisation}
We can pose a measure for this sense of multi-ordered degrees using neighbourhood degree sequences. We could simply divide the number of unique $p$-length sequences by the total number of $p$-length sequences, giving
\begin{equation}\label{p1}
\frac{|\sigma_{p}|}{q_{p}},
\end{equation}
however this is the same no matter how many unique degree sequences occur more than once. Consider the following. Let $c_{pj}$ denote the number of neighbourhood degree sequences of length $p$ in $G$ that have equivalency to $s_{j}\in\sigma_{p}$. Then, for example, take $q_{p} = 5$ and $|\sigma_{p}| = 3$.  We could have $c_{p1} = 1$, $c_{p2} = 1$ and $c_{p3} = 3$ or $c_{p1} = 1$, $c_{p2} = 2$ and $c_{p3} = 2$. Both of these options would have the same value of $\eqref{p1}$, yet the latter has better qualities of being multiply ordered than the former since there are two distinct sequences which occur more than once, rather than just the one in the former case. We can offset \eqref{p1} by considering the differences between the number of $p$-length sequences, $q_{p}$, and the number of occurrences of each (unique) neighbourhood degree sequence in $\sigma_{p}$.  Then $\sum_{j = 1}^{|\sigma_{p}|} c_{pj} = 1$ and we consider the entity
\begin{equation}\label{p2}
\sum_{j = 1}^{|\sigma_{p}|}(q_{p}-c_{pj}).
\end{equation}
This is maximal, $q_{p}(q_{p}-1)$, when all $p$-length neighbourhood degree sequences are unique and zero (i.e. minimal) when all $p$-length neighbourhood degree sequences are equal. We can thus normalise this term as
\begin{equation}
\frac{\sum_{j = 1}^{|\sigma_{p}|}(q_{p}-c_{pj})}{q_{p}(q_{p}-1)}.
\end{equation}
Just taking \eqref{p2} would also not reflect the multi-order requirement. It is really the combination of \eqref{p1} and \eqref{p2} that is required to realise a measure of multi-ordered degrees-- elements of $\sigma_{p}$ should occur frequently and at the same time the number of unique sequences should be as large as possible. Combining \eqref{p1} and \eqref{p2}, then, we get
\begin{equation}
\omega_{p} = \frac{|\sigma_{p}|\sum_{j = 1}^{|\sigma_{p}|}(q_{p}-c_{pj})}{q_{p}^{2}(q_{p}-1)}
\end{equation}
Taking the mean of this over all degrees and subtracting from 1, we have the neighbourhood organisation coefficient
\begin{equation}
\Omega(G) = 1-\frac{1}{|\mathcal{D}_{2}|}\sum_{p=1}^{n-1}\omega_{p}, 
\end{equation}
where $\mathcal{D}_{2}$ is the set of degrees of the graph taken by at least 2 nodes.

\subsection{Updated hierarchical complexity}
Given the above consideration of multi-ordered degrees and the neighbourhood organisation index, we can formulate an update to hierarchical complexity that takes into account multi-ordered degrees. In the terminology of this paper, hierarchical complexity can be written
\begin{equation}\label{origHC}
R(G) = \frac{1}{|\mathcal{D}_{2}|}\sum_{p\in\mathcal{D}_{2}}\frac{1}{p(q_{p}-1)}\sum_{j=1}^{p}\sum_{i\in\mathcal{V}_{p}}(s^{p}_{i}(j)-\mu^{p}(j))^{2}
\end{equation}
where $\mathcal{V}_{p}$ is the set of nodes of degree $p$ and $\mu^{p}(j)$ is the mean of the $j$th entries of all $p$ length neighbourhood degree sequences.

To correct for multi-ordered degrees in this index, we can implement the term $\omega_{p}$ inside the first summand in \label{origHC} to give 
\begin{equation}\label{OHC}
R_{\Omega}(G) = \frac{1}{|\mathcal{D}_{2}|}\sum_{p\in\mathcal{D}_{2}}\frac{\omega_{p}}{p(q_{p}-1)}\sum_{j=1}^{p}\sum_{i\in\mathcal{V}_{p}}(s^{p}_{i}(j)-\mu^{p}(j))^{2}.
\end{equation}
When $\omega_{p}$ is small, multi-orderedness is present in the $p$ degree nodes and thus the value of hierarchical complexity for these degrees is suppressed and vice versa. Computing this for the example in Fig \ref{NhG}.A we obtain $R_{\Omega} = 0.0029$-- a 65 fold decrease from $R$ and a more reasonable expected value of neighbourhood degree sequence diversity.

\subsection{Link to the graph isomorphism problem}
The Weisfeiler-Lehman graph isomorphism test \cite{Weisfeiler1968} is a powerful method for distinguishing labelled graph topologies which holds for almost all graphs \cite{Babai1979}. Based on this test, subtree kernels have been produced for assessing graph similarity in machine learning approaches which are highly efficient compared to other successful kernels \cite{Shervashidze2011}. Indeed, these subtree kernels have been shown to outperform the competition when implemented into a graph neural network approach while mapping similar graph topologies to similar embeddings in a low-dimensional space \cite{Xu2018}.

The subtree of node $i$ of height $h$ constructs a tree rooted at $i$ which extends out to $i$'s neighbours and then out again to $i$'s neighbours' neighbours and so on for $h$ steps, see Fig \ref{NhG}.C. The kernel is a reduction of these subtrees to identifying labels which are then compared between two graphs to check their similarity. Subtrees of height $h=2 \text{ or } 3$ have been shown to achieve best performance in most cases \cite{Shervashidze2011}.

The link to neighbourhood degree sequences then can be established by realising that the information in a subtree of height 2 in an unlabelled graph is completely captured by the node's neighbourhood degree sequence. The length of the neighbourhood degree sequence tells us how many nodes are at height 1 of its subtree kernel (i.e. the degree of the node), while the entries of the sequence tell us how many nodes at height 2 are linked to each node at height 1 (the degrees of each neighbouring node).

\section{Methods}
\subsection{Real-world networks}
Thirty networks were obtained from the network repository \cite{nr2015} from different research domains. Descriptions are kept to a minimum. For further details, we refer the reader to the references.

Social networks: The classical Zachary's karate club network \cite{Zachary1977}, a dolphin social network \cite{Lusseau2003}, the Advogato network \cite{Massa2009}; the anybeat network; the Hamsterster network \cite{Hamsterster}; and a wikivote network \cite{Leskovec2010}.

Biological networks: The macaque cortex network freely available from the BCT was used \cite{Rubinov2010}. This comes as a binary, directed network. To make this undirected we simply took all connections as undirected connections to signify whether or not any connection exists between two regions. We also look at the undirected c.\ elegans metabolic network \cite{Duch2005}; bioGRID protein networks of the fruitfly, mouse and a plant; a yeast protein interaction network \cite{Jeong2001}; and a mouse brain network \cite{Amunts2013}.

Ecological networks: The everglades, florida and mangwet ecosystems networks \cite{Melian2004}.

Economic networks: The global city network is a network of economic ties between cities \cite{Taylor2001}. This is a weighted network which was binarised at 20\% density (20\% of largest weights kept) for our analysis. We also used the beacxc and beaflw economic networks.

Interaction networks: A university email network \cite{Guimera2003}; a Dublin infection network \cite{Infect2012}; and an enron email network \cite{Cohen2005}.

Infrastructure networks: A US and Canada airport network found in the Graph Algorithms in Matlab Code toolbox \cite{USairport}; the euroroad network \cite{Bader2012}; and a grid power network \cite{Watts1998}.

Web networks: the EPA hyperlink network \cite{deNooy2011}; the edu hyperlink network \cite{Gleich2004} and the indochina 2004 hyperlink network \cite{Boldi2011}.

Technological networks: A router network.

In addition, we study a benchmark dataset of 406 real world networks used in \cite{Ghasemian2018} from the Colorado Index of Complex Networks \cite{Clauset2016}. This includes 186 static networks of which just 3 overlap with the above (dolphin social network, Macaque cortex and the uni email network). It also includes two temporal networks relating to the same data of organisation affiliations each with 111 samples taken monthly from May 2002 until August 2011 \cite{Seierstad2011}. The first of these is a network of organisation co-affiliations of directors while the other is a network of co-directorship among organisations.

\subsection{Models}

Configuration models: Random graphs with fixed degree distributions \cite{Maslov2002} were generated using a freely available algorithm in the Brain Connectivity Toolbox \cite{Rubinov2010}. Fifty randomisations were computed for each real world network.

\subsection{Classical global network indices}
Clustering coefficient: The global clustering coefficient, $C$, measures the ratio of closed to open triples in the network. A triple is a path of length two, $\{(i,j),(j,k)\}$, where it is closed if $(k,i)$ also exists in the network and open otherwise. It is a measure of network segregation.

Degree variance: The degree variance, $v = var(k)$, is a measure of network heterogeneity \cite{Snijders1981}. Here we use the normalised version \cite{Smith2018b}.

Characteristic path length: The characteristic path length, $L$, is the average of the shortest paths existing between all pairs of nodes in the network. It is known as a measure of network integration.

Assortativity: Assortativity, $r$, is a correlation of the degrees of nodes which are connected in the network. It is positive if similar degree nodes are generally connected to one another, negative if similar degree nodes are generally not connected to one another and zero if there is no pattern of correlation \cite{Newman2002a}.

Modularity: Modularity, $Q$, measures the propensity of nodes to form into highly connected communities which are less connected to the rest of the network \cite{Newman2004}. 

\section{Experiments}
The supplementary material contains results of indices of a variety of different models-- random graphs \cite{Erdos1959}, random geometric graphs \cite{Dall2002}, small-world models \cite{Watts1998}, scale-free models \cite{Barabasi1999a} and random hierarchy models \cite{Smith2017a}. The main article shall focus on experiments using the most relevant data of all-- over 200 real world networks.

\subsection{Index Correlations}
\begin{figure}[tb]
	\centering
	\includegraphics[trim = 115 55 0 0,clip,scale=.15]{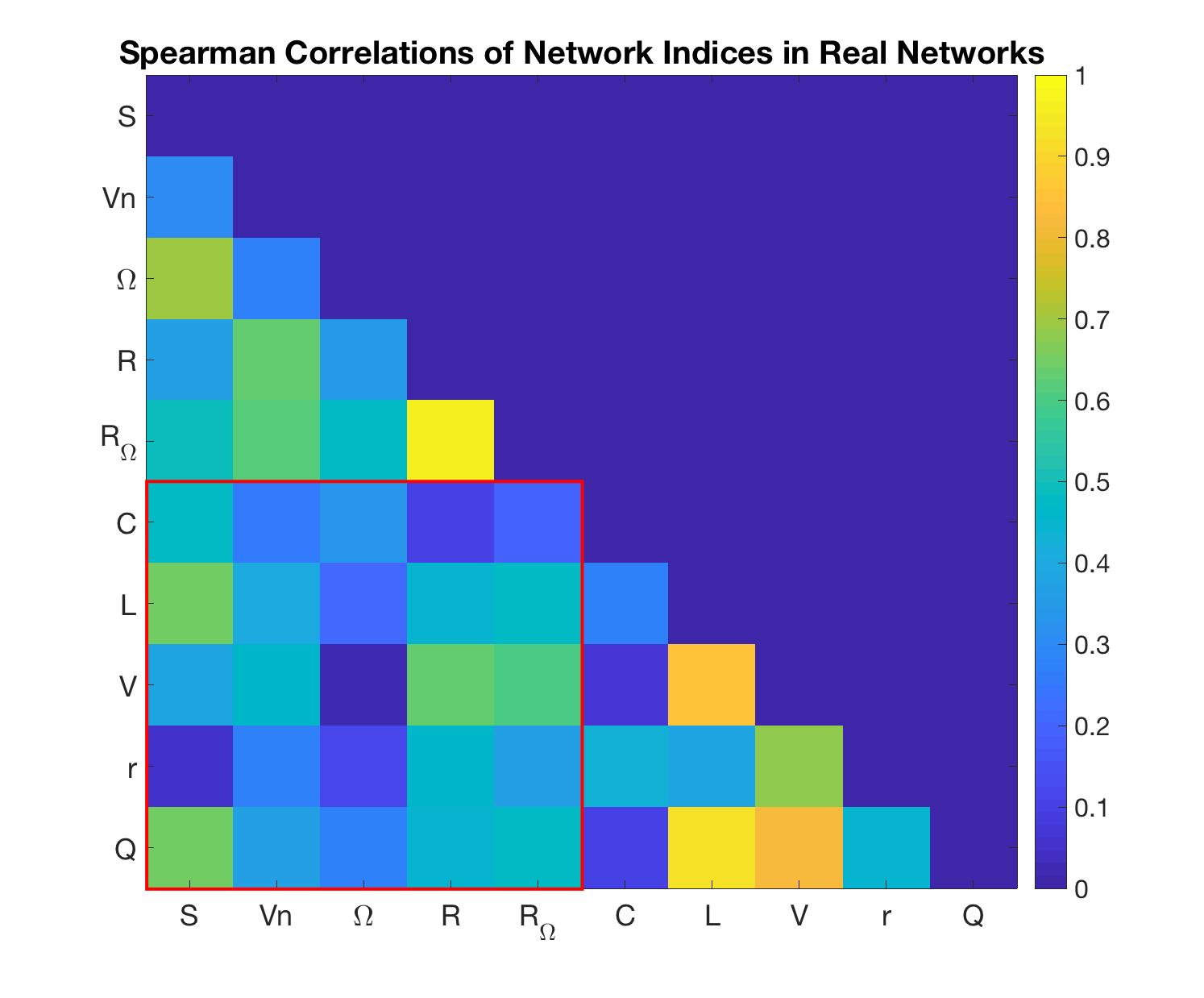}
	\caption{Absolute values of index correlations (Spearman's correlation coefficient) for combined values of small-world, scale-free, random, random geometric, and random hierarchy network models and the thirty real world networks considered (a total of 150 samples). Within the red square are the correlations between neighbourhood degree indices $S$-- neighbourhood similarity, $V_{n}$-- node heterogeneity, $\Omega$-- neighbourhood organisation, $R$-- hierarchical complexity, $R_{\Omega}$-- hierarchical complexity corrected for multi-ordered degrees, and classical network indices $C$- transitivity, $v$-- degree variance, $L$-- characteristic path length, $r$-- assortativity, and $Q$-- modularity.}
	\label{IC}
\end{figure}

Spearman correlations were computed between the proposed indices alongside classical network indices across all real networks, Fig \ref{IC}. We used Spearman's correlation since the values clearly did not follow a normal distribution (i.e. Pearson's correlation would not have been valid). The red box contains all correlations between neighbourhood degree sequence indices and classical network indices. It is clear that there are no observable high correlations between proposed indices and classical indices, providing strong evidence that indeed these new indices explain previously unrealised properties of network topology. Unsurprisingly, $R$ and $R_{\Omega}$ were highly correlated, although the correlation between $\Omega$ and $R_{\Omega}$ was only low to moderate. But the fact there were no strong correlations other than between $R$ and $R_{\Omega}$ ($>0.8$) suggests there is a rich amount of information to be obtained from neighbourhood degree sequences.

On the other hand, among classical network indices, strong correlations were found to exist between the $L$, $V$ and $Q$, indicating that these indices all pointed mostly towards a single topological property of the networks. We suggest that this property is likely to be about the dominance of hub nodes, since these nodes are those which enable general short path lengths, while Newman's modularity is known to be confounded by hubs \cite{Yang2014}.

Although high correlations which are above the standard of 0.8 have been highlighted, there are notable moderate correlations between $L$ and $S$ (0.6477), $Q$ and $S$ (0.6419) and $V$ and $R$ (0.6274). However, the average correlation across all metric pairs has a magnitude of 0.4283, which would be regarded as a low-to moderate correlation. We then have to expect that measurements of a network will likely have some degree of correlation simply due to the fact that they are enacted in measuring the same topologies and since complex networks tend to show broadly consistent features in comparison with random null models. Nonetheless, the standard deviation of the metric correlation magnitudes is 0.2269, putting one standard deviation above the mean at 0.6552 of which none of the moderate correlations previously mentioned lie above. Thus, although in usual terms these are moderate correlations, with respect to complex network metrics they appear to be within reasonable limits to suggest they broadly measure different network properties.

It is also worth recalling that correlation does not mean causation. This means that the general tendency of complex networks to exhibit correlated metrics does not necessarily mean they are measuring the same or similar property in the network, as it may be that networks which have greater modularity have greater characteristic path lengths by virtue of an underlying joint causation.

\subsection{Characteristics of real-world networks}
\begin{figure*}[tb]
	\centering
	\includegraphics[trim = 400 0 0 0,clip,scale=.14]{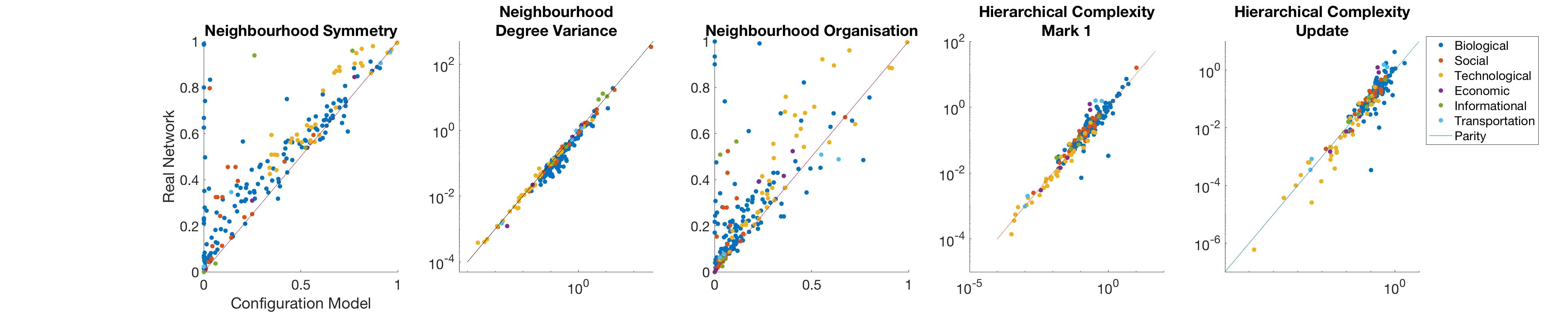}
	\caption{Scatter plots (log-log scale) of real network index values against average values for configuration models. Real world networks have generally well organised neighbourhoods for their degree distributions. They also show a tendency towards hierarchical complexity, although there is a strong drop in statistical significance when taking into account multi-ordered degrees.}
	\label{ex2}
\end{figure*}

All proposed indices were applied to the thirty real-world networks of the Network Repository and the 181 non-overlapping static networks of the ICON, alongside median values taken over the two temporal networks. In addition, ten realisations of configuration models with fixed degree distributions were generated for each real-world network and we compared the neighbourhood indices of the real networks with the average values obtained from configuration models. The results are described for each Network Repository network in Table \ref{sparsebinary}. Scatter plots of all real network values against configuration model values are show in Fig \ref{ex2}.

Although all indices found significant differences between real networks and configuration models, Table \ref{statdifference}, first row, the greatest general differences found were in neighbourhood similarity, $p = 4.74\times10^{-28}$ with a paired ranked effect size of $0.5320$, and in neighbourhood organisation, $p = 1.66\times10^{-23}$ with a paired ranked effect size of 0.4841. This was clearly observed in Fig \ref{ex2}, first and  centre plots, respectively. On the other hand, hierarchical complexity was only weakly greater in real networks than their configuration models. This was even less convincing when we took account of multi-orderedness, increasing the $p$-value to just below 0.05. This is interesting in light of the work done on hierarchical complexity of the human brain function and structure. Hierarchical complexity was not a consistent feature of real world networks and can thus be conjectured as a special feature of brain networks, where a great diversity of functional roles is present \cite{Smith2019}.

\begin{table*}[tbh]
\centering
	\caption{Neighbourhood degree sequence characteristics of 30 real-world networks from the Network Repository. Bracketed values are the means for ten realisations of configuration models. Underneath are the $p$-values for Wilcoxon signed rank tests and their effect sizes between real and edge-randomised values for each index. Legend: s- social, b-biological, tr- transportation, en- economic, in- informational, t- technological. $S$- neighbourhood similarity, $V_{n}$- relative node heterogeneity, $\Omega$- neighbourhood organisation, $R$- hierarchical complexity, $R_{\Omega}$- hierarchical complexity update}
	\resizebox{\textwidth}{!}{
		\begin{tabular}{ccccccccc}
		Type 	& \textbf{Name} 	& $S$ & $V_{n}$ & $\Omega$ & $R$ & $R_{\Omega}$ & \textbf{Size} & \textbf{Density}\\
		\hline
		 & karate club 		&0.324 (0.062)  	&1.714 (2.081)	&0.279 (0.062) 	&0.296 (0.318)	&0.190 (0.269)	& 34		& 0.1390	\\				
		 & hi-tech firm	 	&0.111 (0.044)	&0.811 (0.993)	&0.119 (0.066)	&0.181 (0.183)	&0.176 (0.157)		&36		&0.1444	\\
		 & dolphins	 		&0.113 (0.111)	&0.669 (0.736)	&0.076 (0.062)	&0.045 (0.044) 	&0.040 (0.027)		&62		&0.0841	\\
		 s & wikivote	 		&0.245 (0.249)	&4.299 (4.594)	&0.044 (0.042)	&0.155 (0.145)		&0.139 (0.121)		&889		&0.0074	\\
		 & hamsterer 		&0.455 (0.124)	&3.237 (4.581)	&0.153 (0.012)	&0.218 (0.129)	&0.156 (0.120)		&2426	&0.0057	\\
		 & advogato	 		&0.394 (0.176)	&16.977 (20.231)	&0.016 (0.011)	&0.607 (0.457)	&0.569 (0.420)		&6551	&0.0019	\\
		 & anybeat	 		&0.593 (0.566)	&333.632 (418.489)	&0.017 (0.014)	&15.561 (10.581)	&11.164 (7.276)		&12645	&0.0006	\\	
		 & enron email &0.042 (0.025) &1.657 (1.711) &0.049 (0.019) &0.176 (0.151) &0.168 (0.145)	&143 &0.0614 \\
		& dublin contact &0.015 (0.011) &0.960 (1.137) &0.014 (0.008) &0.074 (0.044) &0.071 (0.044)	 &410 &0.0330 \\
		 & uni email &0.148 (0.144) &1.339 (1.559) &0.029 (0.027) &0.033 (0.027) &0.030 (0.023)	&1133 &0.0085 \\
		\hline
				
		\hline
		 & mouse brain &0.000 (0.000) &0.937 (0.930) &0.630 (0.597) &0.024 (0.016) &0.024 (0.016)	 &213 &0.7160 \\ 
		 & macaque cortex 	&0.050 (0.003)	&1.116 (1.291)	&0.039 (0.003)	&0.436 (0.302)	&0.354 (0.317)		&242 	&0.1047	\\
		 & celegans metabolic	&0.265 (0.015)	&18.957 (17.647)	&0.209 (0.017)	&2.410 (2.513)	&1.757 (2.502)		&453		&0.0198	\\
		b & mouse bioGRID &0.849 (0.792) &4.652 (11.464) &0.271 (0.173) &0.229 (0.206) &0.111 (0.133)	 &1455 &0.0015 \\
		 & plant bioGRID &0.675 (0.579) &1.363 (3.135) &0.122 (0.064) &0.050 (0.027) &0.040 (0.022)	&1745 &0.0020 \\
		 & yeast protein &0.677 (0.664) &2.603 (3.694) &0.153 (0.136) &0.014 (0.016) &0.007 (0.013)	&2114 &0.0010 \\
		 & fruitfly bioGRID &0.404 (0.384) &3.798 (3.775) &0.039 (0.024) &0.022 (0.016) &0.019 (0.014)	 &7282 &0.0009 \\
		 
		& everglades eco &0 (0) &1.300 (1.286) &0.174 (0.178) &0.323 (0.164) &0.323 (0.165)	 &69 &0.3762 \\
		& mangwet eco &0.062 (0) &1.433 (1.323) &0.189 (0.133) &0.402 (0.281) &0.357 (0.239)	 &97 &0.3106 \\
		 & florida  eco &0.070 (0) &1.569 (1.642) &0.087 (0.029) &0.396 (0.192) &0.362 (0.190)	&128 &0.2553 \\	
		\hline

		\hline
		tr & US airports 		&0.022 (0) 	&0.976 (0.834)	&0.076 (0.048)	&1.578 (0.350) 	&1.552 (0.347)		&456 	&0.3658	\\
		 & euroroad &0.906 (0.912) &1.169 (1.327) &0.509 (0.550) &0.001 (0.001) &0.000 (0.000)	 &1174 &0.0021 \\
		\hline
		
		\hline
		 & global cities 		&0.309 (0.249) 	&0.620 (0.606)	&0.417 (0.356)	&0.029 (0.034)	&0.025 (0.035)		&55		&0.2000	\\	
		en & beacxc &0.010 (0.009) &1.332 (1.289) &0.008 (0.008) &1.238 (0.216) &1.238 (0.200)	 &506 &0.332 \\
		 & beaflw &0 (0) &1.316 (1.285) &0 (0) &0.839 (0.226) &0.839 (0.221)	&507 &0.352 \\
		\hline

		\hline
		 & EPA hyperlink &0.958 (0.767) &12.764 (7.723) &0.568 (0.113) &0.054 (0.045) &0.053 (0.070)	 &3031 &0.0014 \\
		in & edu hyperlink &0.538 (0.524) &10.752 (10.903) &0.056 (0.039) &0.073 (0.086) &0.029 (0.031)	 &4772 &0.0008 \\	
		 & indochina hyperlink &0.940 (0.262) &8.608 (5.344) &0.509 (0.029) &0.029 (0.014) &0.016 (0.012)	 &11358 &0.0007 \\
		\hline	
		
		\hline
		t & techrouters 		&0.442 (0.376)	&2.069 (3.254) 	&0.053 (0.028)	&0.077 (0.040)	&0.071 (0.034)	 	&2113	&0.0030	\\
		& power grid &0.862 (0.861) &1.314 (1.514) &0.310 (0.310) &0.001 (0.001) &0.001 (0.000)	 &4941 &0.0005 \\
		
	\end{tabular}\label{sparsebinary}
	}
	\caption{Statistical differences between neighbourhood degree sequence characteristics of 213 real-world networks. Shown are the $p$-values for Wilcoxon signed rank tests and their effect sizes between real and edge-randomised values for each index.}
	\begin{tabular}{ccccccccc}
		& $p$-value &4.74$\times10^{-28}$ &6.07$\times10^{-19}$  &1.66$\times10^{-23}$ &9.25$\times10^{-4}$ &0.048 &  &  \\

		& effect size &0.5320 &-0.4308 &0.4841 &0.1605 &0.0957 &  &  \\

	\end{tabular}\label{statdifference}

\end{table*}

\begin{figure}[tb]
	\centering
	\includegraphics[trim = 45 40 0 0,clip,scale=.24]{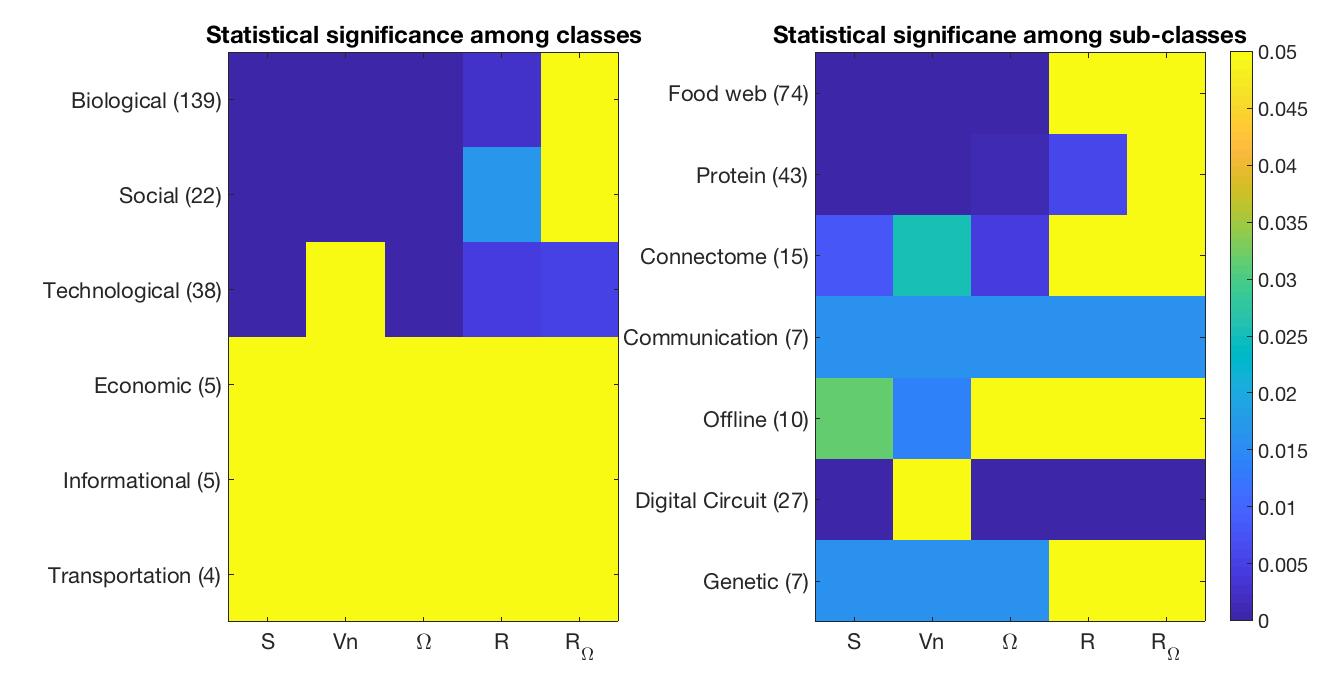}
	\caption{The $p$-values of Wilcoxon sign rank tests of the statistical difference between network indices for classes of real networks and their configuration models.}
	\label{ex3}
\end{figure}

Tentatively, hierarchical complexity also appears to be a strong property of ecological networks. We only studied three such networks here, but all had substantially higher hierarchical complexity than expected for their degree distributions, while other characteristics are not notably different from the expected values, Table \ref{sparsebinary}. 

We then looked at neighbourhood degree sequence properties among different network classes. We applied Wilcoxon sign rank tests, as before, but this time restricted to classes and subclasses of networks, see \cite{Ghasemian2018} for more details. Greater neighbourhood organisation and similarity were found consistently among all classes with a high enough statistical power. On the other hand, technological networks, including digital circuit networks failed to find any difference in neighbourhood heterogeneity between real networks and their configuration models, suggesting a general topological difference between technological networks and biological and social networks, particularly. Interestingly, technological networks (including digital circuit networks) were found to have less hierarchical complexity than their configuration models. We expect that this is to do with a higher degree of order present in digital circuit networks, where different components connect in limited ways, constricted by the logical ordering of electronics. It was also very noticeable that the difference of hierarchical complexity in biological and social networks dropped away when updating for multi-orderedness, suggesting that multi-orderedness is a distinct feature of biological and social networks. In biological networks, this appeared to be driven by protein networks, where food webs and connectomes were not found to be more hierarchically complex than configuration models even from the original definition. The fact that connectomes of animals (3 cat, 5 primate, 2 macaque, 2 nematode, 2 visual cortical neuron level networks in human) were not found to have a general property of hierarchical complexity again suggests the specialness of this feature in the macro-level human brain particularly \cite{Smith2019} and hints towards possible links with intelligence.

\subsection{Neighbourhood organisation in Norwegian director co-affiliation temporal networks}
\begin{figure}[tb]
	\centering
	\includegraphics[trim = 0 10 0 10,clip,scale=.5]{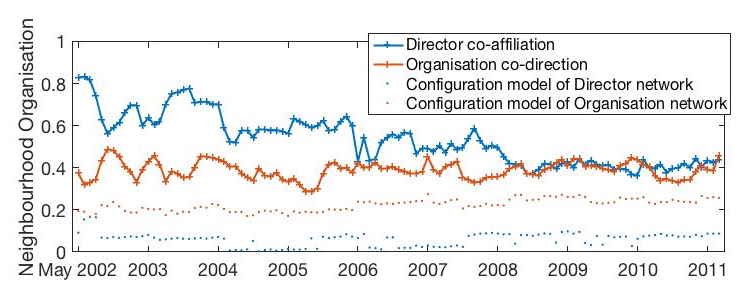}
	\caption{Temporal progression of neighbourhood organisation in the director-organisation affiliation networks}
	\label{ex4}
\end{figure}

In a specific example of revealing new insights into networks using these methods, we undertook an analysis of the two temporal networks included in the ICON corpus. These were monthly sampled social networks of Norwegian company directors, where edges between directors appeared where the two were affiliated with at least one company, and concurrently sampled Norwegian company networks where edges existed where those companies shared a director \cite{Seierstad2011}. Both spanned the same time period from May 2002 to August 2011 and the significance of the data was that, during this time period, legislation was passed to ensure proportional representations of women in directorships to counteract structural inequalities \cite{Seierstad2011}. From an organisational standpoint, it stands to reason that this may cause a fairly dramatic disruption to these networks. Fig \ref{ex4} shows neighbourhood organisation over time for both networks alongside that of their configuration models constructed at each time point.

It is striking that while the company network maintained similar levels of neighbourhood organisation throughout the period, the neighbourhood organisation of the director network steadily decreased throughout the period from roughly 0.8 down to around 0.4 (coinciding with company network levels) by mid 2008 where it stayed until the end of the sampling. No particular trends were notice in either of the configuration models. Looking more closely at the director network trend, it was apparent that the decrease in neighbourhood organisation appeared almost stepwise in two year cycles with steps down around May 2004, 2006 and 2008. This validates the hypothesis that the overhaul in directorships in a short space of time contributed to a substantial disruption to the neighbourhood organisation of the network. Although it is beyond the scope of this study, it would be of interest to seek out explanations for this trend as well as possible correlations with this phenomenon and other factors.

\section{Limitations and future work}
There is significant scope to extend and improve on these proposed methods. A lot of the methods developed here depend on comparing nodes of the same degree, however it would be of great relevance to have this property more relaxed so that comparisons can be done across nodes of similar but not necessarily identical degrees. This is particularly the case for real-world and configuration models where the greater spontaneity of connections means that nodes which exhibit similar properties may differ in degree by one or two connections. Furthermore, this may help to create more reliable indices with less variability within populations.

We demonstrated a link between neighbourhood degree sequences and Weisfeiler-Lehman graph subtree kernels \cite{Shervashidze2011} which provide powerful graph learning results \cite{Xu2018} based on long-standing graph isomorphism results \cite{Weisfeiler1968}. It would be of high interest to undertake a detailed study of the relevance of the neighbourhood degree sequence analyses for interpreting the embedding space of these graph classification approaches as network phenomena. At the same time, this link hints that analysing the diversity and structure of neighbourhood degree sequences within a network-- such as hierarchical complexity and neighbourhood organisation--  is indeed a very powerful and efficient way to describe the topological similarity within a network. Further detailed work is required to substantiate this conjecture.

\section{Conclusion}
We introduced several methods to understand complex networks through neighbourhood degree sequences. These targeted key concepts such as similarity and symmetry, organisation, complexity and heterogeneity. The developed network indices were not found to be strongly correlated with each other nor with classical network indices over 215 real world networks, indicating that neighbourhood degree sequences offer a rich and unique branch of analysis. We found that neighbourhood similarity and neighbourhood organisation were consistent general characteristics of complex networks. Evidence suggested that the hierarchical complexity evident in the human brain was not a general property of animal connectomes. Also, neighbourhood organisation was found to decrease over time in a company director network where the composition of directors went through major alterations, while neighbourhood organisation in the company network remained steady. It is expected that this study will act as a springboard for new methods and applications relating to neighbourhood degree sequences, revealing important insights into networks across various disciplines.

{\footnotesize
\bibliography{References}}

\begin{thebibliography}{10}
\urlstyle{rm}
\expandafter\ifx\csname url\endcsname\relax
  \def\url#1{\texttt{#1}}\fi
\expandafter\ifx\csname urlprefix\endcsname\relax\def\urlprefix{URL }\fi
\expandafter\ifx\csname doiprefix\endcsname\relax\def\doiprefix{DOI: }\fi
\providecommand{\bibinfo}[2]{#2}
\providecommand{\eprint}[2][]{\url{#2}}

\bibitem{Bollobas1981}
\bibinfo{author}{Bollob\'as, B.}
\newblock \bibinfo{journal}{\bibinfo{title}{{Degree sequences of random
  graphs}}}.
\newblock {\emph{\JournalTitle{Discrete Mathematics}}}
  \textbf{\bibinfo{volume}{33}}, \bibinfo{pages}{1--19} (\bibinfo{year}{1981}).

\bibitem{Strogatz2001}
\bibinfo{author}{Strogatz, S.~H.}
\newblock \bibinfo{journal}{\bibinfo{title}{{Exploring complex networks}}}.
\newblock {\emph{\JournalTitle{Nature}}} \textbf{\bibinfo{volume}{410}},
  \bibinfo{pages}{268--276} (\bibinfo{year}{2001}).

\bibitem{Erdos1959}
\bibinfo{author}{{Erd{\"{o}}s P.}} \& \bibinfo{author}{R{\'{e}}nyi, A.}
\newblock \bibinfo{journal}{\bibinfo{title}{{On random graphs}}}.
\newblock {\emph{\JournalTitle{Pubilcationes Mathematicae Debrecen}}}
  \textbf{\bibinfo{volume}{6}}, \bibinfo{pages}{290--297}
  (\bibinfo{year}{1959}).

\bibitem{Dall2002}
\bibinfo{author}{Dall, J.} \& \bibinfo{author}{Christensen, M.}
\newblock \bibinfo{journal}{\bibinfo{title}{{Random geometric graphs}}}.
\newblock {\emph{\JournalTitle{Physical Review E}}}
  \textbf{\bibinfo{volume}{66}}, \bibinfo{pages}{016121}
  (\bibinfo{year}{2002}).

\bibitem{Watts1998}
\bibinfo{author}{{Watts D.J.}} \& \bibinfo{author}{Strogatz, S.~H.}
\newblock \bibinfo{journal}{\bibinfo{title}{{Collective dynamics of small-world
  networks}}}.
\newblock {\emph{\JournalTitle{Nature}}} \textbf{\bibinfo{volume}{393}},
  \bibinfo{pages}{440--442} (\bibinfo{year}{1998}).

\bibitem{Newman2001}
\bibinfo{author}{Newman, M. E.~J.}, \bibinfo{author}{Strogatz, S.~H.} \&
  \bibinfo{author}{J., W.~D.}
\newblock \bibinfo{journal}{\bibinfo{title}{{Random graphs with arbitrary
  degree distributions and their applications}}}.
\newblock {\emph{\JournalTitle{Physical Review E}}}
  \textbf{\bibinfo{volume}{6402}}, \bibinfo{pages}{6118}
  (\bibinfo{year}{2001}).

\bibitem{Maslov2002}
\bibinfo{author}{Maslov, S.} \& \bibinfo{author}{Sneppen, K.}
\newblock \bibinfo{journal}{\bibinfo{title}{{Specificity and Stability in
  Topology of Protein Networks}}}.
\newblock {\emph{\JournalTitle{Science}}} \textbf{\bibinfo{volume}{296}},
  \bibinfo{pages}{910 LP -- 913} (\bibinfo{year}{2002}).

\bibitem{Smith2017a}
\bibinfo{author}{Smith, K.} \& \bibinfo{author}{Escudero, J.}
\newblock \bibinfo{journal}{\bibinfo{title}{{The complex hierarchical topology
  of {EEG} functional connectivity}}}.
\newblock {\emph{\JournalTitle{Journal of Neuroscience Methods}}}
  \textbf{\bibinfo{volume}{276}}, \bibinfo{pages}{1--12}
  (\bibinfo{year}{2017}).

\bibitem{Ravasz2003}
\bibinfo{author}{{Ravasz E.}} \& \bibinfo{author}{Barabasi, A.~L.}
\newblock \bibinfo{journal}{\bibinfo{title}{{Hierarchical organization in
  complex networks}}}.
\newblock {\emph{\JournalTitle{Physical Review E}}}
  \textbf{\bibinfo{volume}{67}}, \bibinfo{pages}{26112} (\bibinfo{year}{2003}).

\bibitem{Kaiser2010}
\bibinfo{author}{{Kaiser M.}}, \bibinfo{author}{{Hilgetag C.C.}} \&
  \bibinfo{author}{K{\"{o}}tter, R.}
\newblock \bibinfo{journal}{\bibinfo{title}{{Hierarchy and dynamics of neural
  networks}}}.
\newblock {\emph{\JournalTitle{Frontiers in Neuroinformatics}}}
  \textbf{\bibinfo{volume}{4}}, \bibinfo{pages}{112} (\bibinfo{year}{2010}).

\bibitem{Barthelemy2004}
\bibinfo{author}{Barth\'elemy, M.}, \bibinfo{author}{Barrat, A.},
  \bibinfo{author}{Pastor-Satorras, R.} \& \bibinfo{author}{Vespignani, A.}
\newblock \bibinfo{journal}{\bibinfo{title}{Velocity and hierarchical spread of
  epidemic outbreaks in scale-free networks}}.
\newblock {\emph{\JournalTitle{Physical Review Letters}}}
  \textbf{\bibinfo{volume}{92}}, \bibinfo{pages}{178701}
  (\bibinfo{year}{2004}).

\bibitem{Smith2017c}
\bibinfo{author}{Smith, K.}, \bibinfo{author}{Ab{\'{a}}solo, D.} \&
  \bibinfo{author}{Escudero, J.}
\newblock \bibinfo{journal}{\bibinfo{title}{{Accounting for the Complex
  Hierarchical Topology of {EEG} Phase-based Functional Connectivity in Network
  Binarisation}}}.
\newblock {\emph{\JournalTitle{PLOS ONE}}} \textbf{\bibinfo{volume}{12}},
  \bibinfo{pages}{e0186164} (\bibinfo{year}{2017}).

\bibitem{Smith2019}
\bibinfo{author}{Smith, K.} \emph{et~al.}
\newblock \bibinfo{journal}{\bibinfo{title}{{Hierarchical Complexity of the
  Adult Human Structural Connectome}}}.
\newblock {\emph{\JournalTitle{Neuroimage}}} \textbf{\bibinfo{volume}{191}},
  \bibinfo{pages}{205--215} (\bibinfo{year}{2019}).

\bibitem{Barrus2018}
\bibinfo{author}{Barrus, M.} \& \bibinfo{author}{Donovan, E.}
\newblock \bibinfo{journal}{\bibinfo{title}{Neighbourhood degree lists of
  graphs}}.
\newblock {\emph{\JournalTitle{Discrete Mathematics}}}
  \textbf{\bibinfo{volume}{341}}, \bibinfo{pages}{175--183}
  (\bibinfo{year}{2018}).

\bibitem{Nishimura2017}
\bibinfo{author}{Nishimura, N.} \& \bibinfo{author}{Subramanya, V.}
\newblock \bibinfo{title}{Graph editing to a given neighbourhood degree list is
  fixed-parameter tractable}.
\newblock In \bibinfo{editor}{Gao, X.}, \bibinfo{editor}{Du, H.} \&
  \bibinfo{editor}{Han, M.} (eds.) \emph{\bibinfo{booktitle}{COCOA 2017:
  Combinatorial optimization and applications}}, vol. \bibinfo{volume}{10628}
  of \emph{\bibinfo{series}{Lecture Notes in Computer Science}},
  \bibinfo{pages}{138--153} (\bibinfo{publisher}{Springer, Cham},
  \bibinfo{year}{2017}).

\bibitem{Bonacich1972}
\bibinfo{author}{{Bonacich P.}}
\newblock \bibinfo{journal}{\bibinfo{title}{{Factoring and weighting approaches
  to clique identification}}}.
\newblock {\emph{\JournalTitle{Journal of Mathematical Sociology}}}
  \textbf{\bibinfo{volume}{2}}, \bibinfo{pages}{113--120}
  (\bibinfo{year}{1972}).

\bibitem{Newman2002}
\bibinfo{author}{Newman, M.}
\newblock \bibinfo{journal}{\bibinfo{title}{{Assortative mixing in networks}}}.
\newblock {\emph{\JournalTitle{Physical Review Letters}}}
  \textbf{\bibinfo{volume}{89}}, \bibinfo{pages}{208701}
  (\bibinfo{year}{2002}).

\bibitem{Sole2004}
\bibinfo{author}{{Sol{\'{e}} R.}} \& \bibinfo{author}{Valverde, S.}
\newblock \bibinfo{title}{{Complex Networks}}.
\newblock vol. \bibinfo{volume}{650} of \emph{\bibinfo{series}{Lecture Notes in
  Physics}}, chap. \bibinfo{chapter}{Informatio}, \bibinfo{pages}{189--207}
  (\bibinfo{publisher}{Springer}, \bibinfo{year}{2004}).

\bibitem{Snijders1981}
\bibinfo{author}{Snijders, T. A.~B.}
\newblock \bibinfo{journal}{\bibinfo{title}{{The degree variance: an index of
  graph heterogeneity}}}.
\newblock {\emph{\JournalTitle{Social Networks}}} \textbf{\bibinfo{volume}{3}},
  \bibinfo{pages}{163--174} (\bibinfo{year}{1981}).

\bibitem{Bell1992}
\bibinfo{author}{Bell, F.~K.}
\newblock \bibinfo{journal}{\bibinfo{title}{{A note on the irregularity of
  graphs}}}.
\newblock {\emph{\JournalTitle{Lin. Alg. Appl.}}}
  \textbf{\bibinfo{volume}{161}}, \bibinfo{pages}{45--64}
  (\bibinfo{year}{1992}).

\bibitem{Estrada2010}
\bibinfo{author}{Estrada, E.}
\newblock \bibinfo{journal}{\bibinfo{title}{{Quantifying network
  heterogeneity}}}.
\newblock {\emph{\JournalTitle{Physical Review E}}}
  \textbf{\bibinfo{volume}{82}}, \bibinfo{pages}{066102}
  (\bibinfo{year}{2010}).

\bibitem{Frucht1939}
\bibinfo{author}{Frucht, R.}
\newblock \bibinfo{journal}{\bibinfo{title}{{Herstellung von Graphen mit
  vorgegebener abstrakter Gruppe}}}.
\newblock {\emph{\JournalTitle{Compositio Mathematica}}}
  \textbf{\bibinfo{volume}{6}}, \bibinfo{pages}{239--250}
  (\bibinfo{year}{1939}).

\bibitem{Weisfeiler1968}
\bibinfo{author}{Weisfeiler, B.} \& \bibinfo{author}{Lehman, A.}
\newblock \bibinfo{journal}{\bibinfo{title}{A reduction of a graph to a
  canonical form and an algebra arising during this reduction}}.
\newblock {\emph{\JournalTitle{Nauchno-Technicheskaya Informatsiya}}}
  \textbf{\bibinfo{volume}{2}}, \bibinfo{pages}{12--16} (\bibinfo{year}{1968}).

\bibitem{Babai1979}
\bibinfo{author}{Babai, L.} \& \bibinfo{author}{Kucera, L.}
\newblock \bibinfo{title}{Canonical labelling of graphs in linear average
  time}.
\newblock In \emph{\bibinfo{booktitle}{Proceedings Symposium on Foundations of
  Computer Science}}, \bibinfo{pages}{39--46} (\bibinfo{year}{1979}).

\bibitem{Shervashidze2011}
\bibinfo{author}{Shervashidze, N.}, \bibinfo{author}{Schweitzer, P.},
  \bibinfo{author}{van Leeuwen, E.}, \bibinfo{author}{Mehlhorn, K.} \&
  \bibinfo{author}{Borgwardt, K.}
\newblock \bibinfo{journal}{\bibinfo{title}{Weisfeiler-lehman graph kernels}}.
\newblock {\emph{\JournalTitle{Journal of Machine Learning Research}}}
  \textbf{\bibinfo{volume}{12}}, \bibinfo{pages}{2539--2561}
  (\bibinfo{year}{2011}).

\bibitem{Xu2018}
\bibinfo{author}{Xu, K.}, \bibinfo{author}{Hu, W.}, \bibinfo{author}{Leskovec,
  J.} \& \bibinfo{author}{Jegelka, S.}
\newblock \bibinfo{title}{How powerful are graph neural networks?}
  (\bibinfo{year}{2018}).
\newblock \bibinfo{note}{Https://arxiv.org/abs/1810.00826}.

\bibitem{nr2015}
\bibinfo{author}{Rossi, R.~A.} \& \bibinfo{author}{Ahmed, N.~K.}
\newblock \bibinfo{title}{The network data repository with interactive graph
  analytics and visualization}.
\newblock In \emph{\bibinfo{booktitle}{Proceedings of the Twenty-Ninth AAAI
  Conference on Artificial Intelligence}} (\bibinfo{year}{2015}).

\bibitem{Zachary1977}
\bibinfo{author}{Zachary, W.~W.}
\newblock \bibinfo{journal}{\bibinfo{title}{{An Information Flow Model for
  Conflict and Fission in Small Groups}}}.
\newblock {\emph{\JournalTitle{J. Anthro. Research}}}
  \textbf{\bibinfo{volume}{33}}, \bibinfo{pages}{452--473}
  (\bibinfo{year}{1977}).

\bibitem{Lusseau2003}
\bibinfo{author}{Lusseau, D.} \emph{et~al.}
\newblock \bibinfo{journal}{\bibinfo{title}{The bottlenose dolphin community of
  doubtful sound features a large proportion of long-lasting associations}}.
\newblock {\emph{\JournalTitle{Behavioral Ecology and Sociobiology}}}
  \textbf{\bibinfo{volume}{54}}, \bibinfo{pages}{396--405}
  (\bibinfo{year}{2003}).

\bibitem{Massa2009}
\bibinfo{author}{Massa, P.}, \bibinfo{author}{Salvetti, M.} \&
  \bibinfo{author}{Tomasoni, D.}
\newblock \bibinfo{title}{Bowling alone and trust decline in social network
  sites}.
\newblock In \emph{\bibinfo{booktitle}{Dependable, Autonomic and Secure
  Computing, 2009. DASC'09. Eighth IEEE International Conference on}},
  \bibinfo{pages}{658--663} (\bibinfo{organization}{IEEE},
  \bibinfo{year}{2009}).

\bibitem{Hamsterster}
\bibinfo{author}{Hamsterster}.
\newblock \bibinfo{title}{Hamsterster social network}.
\newblock \bibinfo{note}{Http://www.hamsterster.com}.

\bibitem{Leskovec2010}
\bibinfo{author}{Leskovec, J.}, \bibinfo{author}{Huttenlocher, D.} \&
  \bibinfo{author}{Kleinberg, J.}
\newblock \bibinfo{title}{Signed networks in social media}.
\newblock In \emph{\bibinfo{booktitle}{Proceedings of the SIGCHI Conference on
  Human Factors in Computing Systems}}, \bibinfo{pages}{1361--1370}
  (\bibinfo{organization}{ACM}, \bibinfo{year}{2010}).

\bibitem{Rubinov2010}
\bibinfo{author}{{Rubinov M.}} \& \bibinfo{author}{Sporns, O.}
\newblock \bibinfo{journal}{\bibinfo{title}{{Complex network measures of brain
  connectivity: uses and interpretations}}}.
\newblock {\emph{\JournalTitle{NeuroImage}}} \textbf{\bibinfo{volume}{52}},
  \bibinfo{pages}{1059--1069} (\bibinfo{year}{2010}).

\bibitem{Duch2005}
\bibinfo{author}{Duch, J.} \& \bibinfo{author}{Arenas, A.}
\newblock \bibinfo{journal}{\bibinfo{title}{Community identification using
  extremal optimization phys}}.
\newblock {\emph{\JournalTitle{Rev. E}}} \textbf{\bibinfo{volume}{72}},
  \bibinfo{pages}{027104} (\bibinfo{year}{2005}).

\bibitem{Jeong2001}
\bibinfo{author}{Jeong, H.}, \bibinfo{author}{Mason, S.},
  \bibinfo{author}{Barabasi, A.} \& \bibinfo{author}{Oltvai, Z.}
\newblock \bibinfo{journal}{\bibinfo{title}{Lethality and centrality in protein
  networks}}.
\newblock {\emph{\JournalTitle{arXiv preprint cond-mat/0105306}}}
  (\bibinfo{year}{2001}).

\bibitem{Amunts2013}
\bibinfo{author}{Amunts, K.} \emph{et~al.}
\newblock \bibinfo{journal}{\bibinfo{title}{Bigbrain: An ultrahigh-resolution
  3d human brain model}}.
\newblock {\emph{\JournalTitle{Science}}} \textbf{\bibinfo{volume}{340}},
  \bibinfo{pages}{1472--1475} (\bibinfo{year}{2013}).

\bibitem{Melian2004}
\bibinfo{author}{Meli{\'a}n, C.~J.} \& \bibinfo{author}{Bascompte, J.}
\newblock \bibinfo{journal}{\bibinfo{title}{Food web cohesion}}.
\newblock {\emph{\JournalTitle{Ecology}}} \textbf{\bibinfo{volume}{85}},
  \bibinfo{pages}{352--358} (\bibinfo{year}{2004}).

\bibitem{Taylor2001}
\bibinfo{author}{Taylor, P.}
\newblock \bibinfo{journal}{\bibinfo{title}{Specification of the world city
  network}}.
\newblock {\emph{\JournalTitle{Geographical Analysis}}}
  \textbf{\bibinfo{volume}{33}}, \bibinfo{pages}{181--194}
  (\bibinfo{year}{2001}).

\bibitem{Guimera2003}
\bibinfo{author}{Guimera, R.}, \bibinfo{author}{Danon, L.},
  \bibinfo{author}{Diaz-Guilera, A.}, \bibinfo{author}{Giralt, F.} \&
  \bibinfo{author}{Arenas, A.}
\newblock \bibinfo{journal}{\bibinfo{title}{Self-similar community structure in
  a network of human interactions}}.
\newblock {\emph{\JournalTitle{Physical review E}}}
  \textbf{\bibinfo{volume}{68}}, \bibinfo{pages}{065103}
  (\bibinfo{year}{2003}).

\bibitem{Infect2012}
\bibinfo{author}{{SocioPatterns}}.
\newblock \bibinfo{title}{Infectious contact networks}.
\newblock \bibinfo{note}{Http://www.sociopatterns.org/datasets/. Accessed
  09/12/12.}

\bibitem{Cohen2005}
\bibinfo{author}{Cohen, W.}
\newblock \bibinfo{title}{Enron email dataset}.
\newblock \bibinfo{note}{Http://www.cs.cmu.edu/~enron/. Accessed in 2009.}

\bibitem{USairport}
\bibinfo{title}{{The US airport network}}.
\newblock
  \bibinfo{howpublished}{https://www.mathworks.com/matlabcentral/mlc-downloads/downloads/submissions/24134/versions/1/
  previews/gaimc/demo/html/airports.html?access{\_}key=}.

\bibitem{Bader2012}
\bibinfo{author}{Bader, D.~A.}, \bibinfo{author}{Meyerhenke, H.},
  \bibinfo{author}{Sanders, P.} \& \bibinfo{author}{Wagner, D.}
\newblock \bibinfo{title}{Graph partitioning and graph clustering}.
\newblock In \emph{\bibinfo{booktitle}{10th DIMACS Implementation Challenge
  Workshop}} (\bibinfo{year}{2012}).

\bibitem{deNooy2011}
\bibinfo{author}{De~Nooy, W.}, \bibinfo{author}{Mrvar, A.} \&
  \bibinfo{author}{Batagelj, V.}
\newblock \emph{\bibinfo{title}{Exploratory social network analysis with
  Pajek}}, vol.~\bibinfo{volume}{27} (\bibinfo{publisher}{Cambridge University
  Press}, \bibinfo{year}{2011}).

\bibitem{Gleich2004}
\bibinfo{author}{Gleich, D.}, \bibinfo{author}{Zhukov, L.} \&
  \bibinfo{author}{Berkhin, P.}
\newblock \bibinfo{journal}{\bibinfo{title}{Fast parallel pagerank: A linear
  system approach}}.
\newblock {\emph{\JournalTitle{Yahoo! Research Technical Report YRL-2004-038}}}
  \textbf{\bibinfo{volume}{13}}, \bibinfo{pages}{22} (\bibinfo{year}{2004}).

\bibitem{Boldi2011}
\bibinfo{author}{Boldi, P.}, \bibinfo{author}{Rosa, M.},
  \bibinfo{author}{Santini, M.} \& \bibinfo{author}{Vigna, S.}
\newblock \bibinfo{title}{Layered label propagation: A multiresolution
  coordinate-free ordering for compressing social networks}.
\newblock In \emph{\bibinfo{booktitle}{WWW}}, \bibinfo{pages}{587--596}
  (\bibinfo{year}{2011}).

\bibitem{Ghasemian2018}
\bibinfo{author}{Ghasemian, A.}, \bibinfo{author}{Hosseinmardi, H.} \&
  \bibinfo{author}{Clauset, A.}
\newblock \bibinfo{title}{Evaluating overfit and underfit in models of network
  community structure}.
\newblock \bibinfo{howpublished}{https://arxiv.org/abs/1802.10582}.

\bibitem{Clauset2016}
\bibinfo{author}{Clauset, A.}, \bibinfo{author}{Tucker, E.} \&
  \bibinfo{author}{Sainz, M.}
\newblock \bibinfo{title}{The colorado index of complex networks}.

\bibitem{Seierstad2011}
\bibinfo{author}{Seierstad, C.} \& \bibinfo{author}{Opsahl, T.}
\newblock \bibinfo{journal}{\bibinfo{title}{For the few not the many? the
  effects of affirmative action on presence, prominence, and social capital of
  women directors in norway}}.
\newblock {\emph{\JournalTitle{Scandinavian Journal of Management}}}
  \textbf{\bibinfo{volume}{27}}, \bibinfo{pages}{44--54}
  (\bibinfo{year}{2011}).

\bibitem{Barabasi1999a}
\bibinfo{author}{Barab{\'{a}}si, A.-L.} \& \bibinfo{author}{Albert, R.}
\newblock \bibinfo{journal}{\bibinfo{title}{{Emergence of Scaling in Random
  Networks}}}.
\newblock {\emph{\JournalTitle{Science}}} \textbf{\bibinfo{volume}{286}},
  \bibinfo{pages}{509 LP -- 512} (\bibinfo{year}{1999}).

\bibitem{Smith2018b}
\bibinfo{author}{Smith, K.} \& \bibinfo{author}{Escudero, J.}
\newblock \bibinfo{title}{Normalised degree variance}.
\newblock \bibinfo{note}{Https://arxiv.org/abs/1803.03057}.

\bibitem{Newman2002a}
\bibinfo{author}{Newman, M.}
\newblock \bibinfo{journal}{\bibinfo{title}{{Assortative mixing in networks}}}.
\newblock {\emph{\JournalTitle{Physical Review Letters}}}
  \textbf{\bibinfo{volume}{89}}, \bibinfo{pages}{208701}
  (\bibinfo{year}{2002}).

\bibitem{Newman2004}
\bibinfo{author}{{Newman M.E.J.}} \& \bibinfo{author}{Girvan, M.}
\newblock \bibinfo{journal}{\bibinfo{title}{{Finding and evaluating community
  structure in networks}}}.
\newblock {\emph{\JournalTitle{Physical Review E}}}
  \textbf{\bibinfo{volume}{69}}, \bibinfo{pages}{26113} (\bibinfo{year}{2004}).

\bibitem{Yang2014}
\bibinfo{author}{Yang, J.} \& \bibinfo{author}{Leskovec, J.}
\newblock \bibinfo{journal}{\bibinfo{title}{Overlapping communities explain
  core-periphery organization of networks}}.
\newblock {\emph{\JournalTitle{Proceedings of the IEEE}}}
  \textbf{\bibinfo{volume}{102}}, \bibinfo{pages}{1892--1902}
  (\bibinfo{year}{2014}).

\end{thebibliography}

\section{Acknowledgements}
We would like to thank Aaron Clauset for helpful discussions and provision of the data from the Colorado Index of Complex Networks. This work was supported by Health Data Research UK (MRC ref Mr/S004122/1), which is funded by the UK Medical Research Council, Engineering and Physical Sciences Research Council, Economic and Social Research Council, National Institute for Health Research (England), Chief Scientist Office of the Scottish Government Health and Social Care Directorates, Health and Social Care Research and Development Division (Welsh Government), Public Health Agency (Northern Ireland), British Heart Foundation and Wellcome. A version of this article has been made available on an online preprint server at https://arxiv.org/abs/1901.02353.

\section{Competing interests}
The author declares no competing interests.

\section{Author Contributions}
KS is the sole author and did all the work.

\section{Data Availability}
The real data used in the manuscript were obtained freely online as noted in section III.A. Code for computing the network models and novel indices are available on the Open Science Framework at doi: 10.17605/OSF.IO/W7BK6.

\end{document}